\documentclass[12pt]{iopart}
\usepackage{float}
\usepackage[dvipdfm]{graphicx}
\usepackage{epsfig}
\usepackage{bm}
\usepackage[flushleft]{threeparttable}
\usepackage[usenames, dvipsnames]{color}
\newcommand{\beq}{\begin{eqnarray}}
\newcommand{\eeq}{\end{eqnarray}}  
\begin{document}

\title[]{Controlling phase transition in monolayer metal diiodides XI$_{2}$ (X: Fe, Co, and Ni) by carrier doping}

\author{Teguh Budi Prayitno}
 
\address{Department of Physics, Faculty of Mathematics and Natural Science, Universitas Negeri Jakarta, Kampus A Jl. Rawamangun Muka, Jakarta Timur 13220, Indonesia}
\ead{teguh-budi@unj.ac.id}
\vspace{10pt}

\begin{abstract}
We applied the generalized Bloch theorem to verify the ground state (most stable state) in monolayer metal diiodides 1T-XI$_{2}$ (X: Fe, Co, and Ni), a family of metal dihalides, using the first-principles calculations. The ground state, which can be ferromagnetic, antiferromagnetic, or spiral state, was specified by a wavevector in the primitive unit cell. While the ground state of FeI$_{2}$ is ferromagnetic, the spiral state becomes the ground state for CoI$_{2}$ and NiI$_{2}$. Since the multiferroic behavior in the metal dihalide can be preserved by the spiral structure, we believe that CoI$_{2}$ and NiI$_{2}$ are promising multiferroic materials in the most stable state. When the lattice parameter increases, we also show that the ground state of NiI$_{2}$ changes to a ferromagnetic state while others still keep their initial ground states. For the last discussion, we revealed the phase transition manipulated by hole-electron doping due to the spin-spin competition between the ferromagnetic superexchange and the antiferromagnetic direct exchange. These results convince us that metal diiodides have many benefits for future spintronic devices.             
\end{abstract}

\vspace{2pc}

\section{Introduction}
 
Recently, multiferroic properties in materials bring great attention from point of view of both theoretical aspects and experimental researches \cite{Tokura1, Cheong, Tokura2}. These properties happen due to a combination of the magnetic and dielectric orders appearing in the same phase. It then leads to extraordinary applications in spintronics especially for the data storage devices \cite{Spaldin, Ramesh, Scott}. However, coupling between those two orders to generate multiferroic behavior, in general, is very weak. A possible resolution to encounter this problem is to introduce the magnetically-induced ferroelectricity in some frustrated helimagnets which leads to the so-called giant magnetoelectric properties \cite{Zhai}.        

Previous authors reported that the multiferroicity in some layered metal dihalides may appear in the spiral (SP) structure \cite{Tokunaga, Wu}. This is triggered by the ferroelectric polarization induced by the magnetic SP structure, thus combining the ferroelectricity and magnetism in the same state. However, there are no sufficient reports if monolayer two-dimensional metal dihalides may also possess the SP ground state. Former authors only tabulated the ferromagnetic (FM) state or the antiferromagnetic (AFM) state as the most stable state and excluded the SP ground state \cite{Kulish, Botana}. It can be understood since exploring the SP state needs a very large unit cell, especially for a long period, thus yielding high computational cost. The SP state itself is a typical helimagnetic state with a fixed cone angle \cite{Kurz}. Some applications of the SP structure can be seen in constructing the domain wall \cite{Sabirianov, Pyatakov, Chen}.

Metal dihalides are layered compounds due to a weak van der Waals (vdW) interaction between layers. In the two-dimensional system, the magnetic order in these materials ignores the Mermin-Wagner theorem which prohibits magnetism due to a strong suppression from the thermal fluctuation. This happens because the vdW type yields the magnetic anisotropy to keep a long-range magnetic order that overcomes the thermal fluctuations, thus leading to the intrinsic ferromagnetism \cite{Wang, Han}. Therefore, it is believed that the two-dimensional metal dihalides are more prominent than any other two-dimensional materials such as graphene and MoS$_{2}$, which have no intrinsic ferromagnetism. Besides, we are also allowed to exfoliate the layered structure into a stable monolayer structure \cite{McGuire2}. In the monolayer limit, the magnetism in the metal dihalides can be tuned either by strain \cite{Mushtaq} or by Hubbard $U$ \cite{Li}. It has also been reported that the magnetism in the monolayer metal dihalides exhibits a topological insulator \cite{Chen1} and a half-metal \cite{Ashton}.               

Our purpose is to explore the SP ground state in the monolayer metal diiodides XI$_{2}$ (X: Fe, Co, and Ni), one of the families of metal dihalides, by utilizing the generalized Bloch theorem (GBT), which is widely used to investigate the SP ground state in some materials \cite{Knopfle, Sandratskii1, Garcia, Kunes}, to minimize the high computational cost. The interesting case of metal diiodides, CoI$_{2}$ and NiI$_{2}$ exhibit multiferroicity for the layered structures \cite{Kurumaji1, Kurumaji2}. In this case, we use a primitive unit cell that consists of a 3$d$ transition magnetic atom and two non-magnetic I atoms. We set a flat spiral formation instead of a conical spiral to include the FM, SP, and AFM states as the most stable states. These states will be assigned by a spiral vector from the rotation of the magnetic moment of the magnetic atom. 

We confirm that the ground state of CoI$_{2}$ and NiI$_{2}$ is an SP ground state, in good suitability with the layered systems \cite{Kurumaji1, Kurumaji2}. Meanwhile, we show an FM ground state in FeI$_{2}$, which also confirms a half-metallic behavior as reported in Ref. \cite{Kovaleva}. When we enhance the lattice parameter, the SP ground state of NiI$_{2}$ changes to the FM ground state, thus showing a sensitivity of state to the large lattice parameters. Meanwhile, FeI$_{2}$ and CoI$_{2}$ still preserve their ground states, showing the robustness of state to the lattice parameters. However, the magnetic moment for all the systems seems robust to the lattice parameter. Based on these results, we assure that multiferroicity in the metal dihalides should also occur in the monolayer systems.

For the last session, we display the dependence of ground state on hole-electron doping, which reveals the phase transition of state in the certain interval of doping. Since the phase transition depends on the competition between the FM superexchange and AFM direct exchange, we show a simple mechanism of how the phase transition occurs. The interesting case is that the FM ground state can be changed to the SP ground state by applying hole doping. Therefore, we believe that introducing doping may also achieve the multiferroic properties in the monolayer metal dihalides.                    
       
\section{Computational Details}
\begin{figure}[h!]
\vspace{-2mm}
\quad\quad\includegraphics[scale=0.7, width =!, height =!]{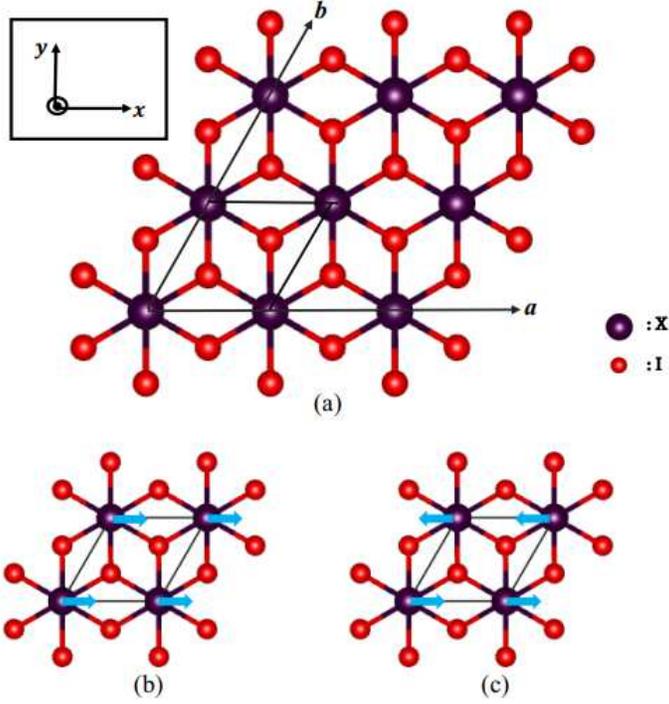}
\vspace{2mm}
\caption{\label{model}(Color online) Crystal structure of monolayer 1T-XI$_{2}$ from top view (a). Here, the FM and AFM ground states are achieved by assigning $\phi=0$ (b) and $\phi=1$ (c), respectively.} 
\end{figure}

The crystal structure of monolayer 1T-XI$_{2}$ (X: Fe, Co, and Ni) is depicted by Fig. \ref{model}(a). The primitive unit cell shown by a parallelogram contains a transition metal X atom (cation) and two I atoms (anion). In the bulk layered structures, FeI$_{2}$ and CoI$_{2}$ crystallize into a CdI$_{2}$ structure ($P\overline{3}m1$), but CoI$_{2}$ does into a CdCl$_{2}$ structure ($R\overline{3}m$) \cite{McGuire}. To avoid the layer-layer interactions in the $z$- axis, we assigned a vacuum distance of more than 17 {\AA}. Meanwhile, we specified the experimental lattice parameters of 4.04 {\AA} for FeI$_{2}$, 3.96 {\AA} for CoI$_{2}$, and 3.89 {\AA} for NiI$_{2}$ \cite{Kulish}. In the periodic lattices, we established the lattice vectors 
\beq
\textit{\textbf{a}}=a\hat{e}_{x}, \qquad  \textit{\textbf{b}}=\frac{a}{2}\hat{e}_{x}+\frac{a}{2}\sqrt{3}\hat{e}_{y},
\eeq 
and the corresponding reciprocal lattice vectors
\beq
\textit{\textbf{A}}=\frac{2\pi}{a}\hat{e}_{x}-\frac{2\pi}{a\sqrt{3}}\hat{e}_{y}, \qquad  \textit{\textbf{B}}=\frac{4\pi}{a\sqrt{3}}\hat{e}_{y},
\eeq 
where $a$ defines the lattice parameter.

The computations were carried out in a $20 \times 20 \times 1$ $k$-point sampling by a linear combination of pseudo-atomic orbitals (LCPAO) approach \cite{Ozaki1, Ozaki2} combined with the norm-conserving pseudopotential \cite{Troullier} as implemented in the OpenMX code \cite{Openmx}. For treating the electron-electron interactions, we selected the generalized gradient approximation (GGA) according to Perdew, Burke, and Ernzerhof \cite{Perdew} with an energy cutoff of 200 Ryd. For the basis sets, we used an abbreviation s$n$p$m$d$a$f$c$, which means that s, p, d, f are the primitive orbitals and $n, m, a, c$ are the number of orbitals. We set s3p3d3 for Fe atom, s3p3d3f2 for Co atom, s3p3d2f1 for Ni atom, and s2p2d1 for I atom. In addition, the cutoff radii as the confinement at the atomic site, beyond which the orbitals vanish, are 4.0 Bohr for X atom and 7.0 Bohr for I atom. The choices of basis sets and cutoff radii are the minimum requirements to get converged results with good accuracy.     

To use the GBT via density functional theory (DFT), we introduced a wavevector $\mathbf{q}$ into an LCPAO \cite{Teguh} 
\beq
\psi_{\mu\mathbf{k}}\left(\mathbf{r}\right)&=&\frac{1}{\sqrt{N}}\left[\sum_{n}^{N}e^{i(\mathbf{k}-\mathbf{q}/2)\cdot\mathbf{R}_{n}}\sum_{i\alpha}C_{\mu\mathbf{k},i\alpha}^{\uparrow}\xi_{i\alpha}\left(\mathrm{\mathbf{r}-\tau_{i}-\mathbf{R}_{n}}\right)
\left(
\begin{array}{cc}
1\\
0\end{array} 
\right)\right.\nonumber\\
& &\left.+\sum_{n}^{N}e^{i(\mathbf{k}+\mathbf{q}/2)\cdot\mathbf{R}_{n}}\sum_{i\alpha}C_{\mu\mathbf{k},i\alpha}^{\downarrow}\xi_{i\alpha}\left(\mathrm{\mathbf{r}-\tau_{i}-\mathbf{R}_{n}}\right)\left(
\begin{array}{cc}
0\\
1\end{array}
\right)\right],\label{lcpao}
\eeq  
where $\xi_{i\alpha}$ describes the localized orbital. To create the spiral configurations, the magnetic moment of metal atom is rotated from site to site obeying \cite{Sandratskii2}
\beq
	\mathbf{M}_{i}=M_{i} \left(
\begin{array}{cc}
\cos\left(\varphi_{i}^{0}+\mathbf{q}\cdot \mathbf{R}_{i}\right)\sin\theta_{i}\\
\sin\left(\varphi_{i}^{0}+\mathbf{q}\cdot \mathbf{R}_{i}\right)\sin\theta_{i}\\
\cos\theta_{i}\end{array} 
\right). \label{moment}  
\eeq
From Eq. \ref{moment}, we see that the cone angle $\theta$ should be constant while the azimuthal angle $\varphi$ will be rotated along $\mathbf{q}$ direction. Since we want to tune the FM, SP, and AFM states by $\mathbf{q}$, we apply the flat spiral by setting the initial direction of magnetic moment of metal atom to $\theta=\pi/2$. Here, we concern the direction of $\mathbf{q}$ along $x$- axis, which is given by 
\beq  
\textit{\textbf{q}}=\phi(\textit{\textbf{A}}+0.5\textit{\textbf{B}})=\phi\frac{2\pi}{a}\hat{e}_{x},
\eeq 
where $\phi=0$ and $\phi=1$ give the FM state (Fig. \ref{model}(b)) and AFM state (Fig. \ref{model}(c)), respectively. Then, the SP state will be generated by specifying $\phi$ between 0 and 1. We also applied the penalty functional to keep $\theta=\pi/2$. 

\section{Results and Discussions}
For the convenience, we divide this section into two subsections, i.e., the non-doped and doped cases
\subsection{Non-doped Case}
\begin{figure}[h!]
\vspace{2mm}
\quad\quad\includegraphics[scale=0.5, width =!, height =!]{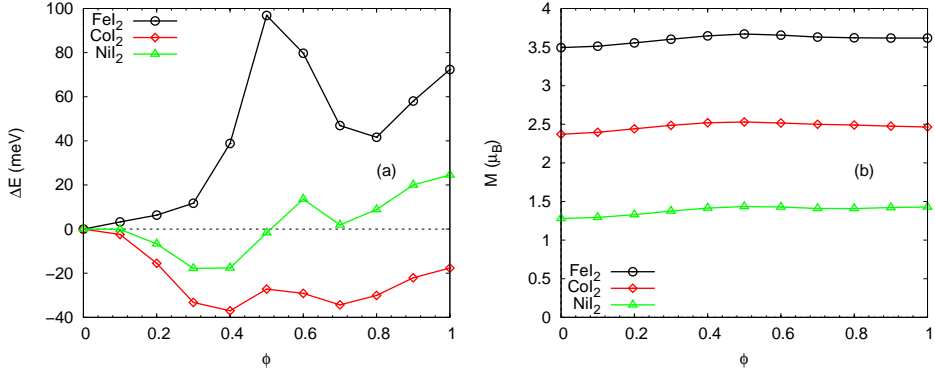}
\vspace{4mm}
\caption{\label{stiff} (Color online) Ground states of some metal diiodides (a) and related magnetic moments (b) for the non-doped case. Here, the ground state is associated with the minimum energy $\Delta E=E(\phi)-E(\phi=0)$.} 
\end{figure}

Figure \ref{stiff} shows the ground states of XI$_{2}$ (X: Fe, Co, and Ni) and magnetic moments for the non-doped case. Here, the ground state is determined by the lowest total energy difference $\Delta E$ as a function of $\phi$. We observe that the ground state of FeI$_{2}$ is an FM state while the ground state of CoI$_{2}$ and NiI$_{2}$ is an SP state. As for CoI$_{2}$ and NiI$_{2}$, our SP ground states are not predicted by Refs. \cite{Kulish, Botana}, where they only considered the FM and AFM states. By observing Fig. \ref{stiff}(a), we deduce that the AFM (FM) state is more stable than the FM (AFM) state for CoI$_{2}$ (NiI$_{2}$), in good agreement with Refs. \cite{Kulish, Botana}. Detailed data can be found in table 1. Moreover, we also notice that all the magnetic moment remains stable with respect to $\phi$, starting from the higher magnetic moment in FeI$_{2}$ and then gradually reducing to NiI$_{2}$. This tendency is in nice agreement with Hund's rule for the free transition metal atoms. 

The appearing SP state without any external treatments exhibits a frustrated spin as the FM and AFM orders compete with each other. This competition of magnetic orders will lead to one of stable ground states, namely, the FM, AFM, or SP state. A similar stable SP state also occurs in the monoatomic transition metal chain and square lattice, which is caused by the spin-spin interaction between the metal atom and its nearest neighbor \cite{Tows}. They also showed that the ground state is influenced by the distance of nearest-neighbor atoms having different configurations of magnetic moment, for example, the FM and AFM order. Similar discussions on the frustrated spin due to spin-spin interaction can also be found in Refs. \cite{Zeleny, Saubanere}. In the other situations, some materials require external treatments to produce the SP state, either using the doping \cite{Inoue} or electric field \cite{Teguh3}, leading to the new stable form. Nevertheless, these treatments usually reduce the magnetic moment, thus decreasing the magnetism in the magnetic materials.    
\begin{table}[ht]
\vspace{2 mm}
\caption{Predicted ground states and magnetic moments of metal diiodides from some DFT calculations for the non-doped case. Here, $GS$ and $M$ denote the ground state and magnetic moment in units of $\mu_{B}$, respectively.}   
\centering 
\begin{tabular}{c c c c c} 
\hline 
Metal&$(GS, M)^{1}$&$(GS, M)^{2}$&$(GS, M)^{3}$\\ 
\hline 
FeI$_{2}$&(FM, 3.49)&(FM, 4.0)&(FM, 3.45)\\
CoI$_{2}$&(SP, 2.52)&(AFM, 3.0)&(AFM, 2.23)\\
NiI$_{2}$&(SP, 1.38)&(FM, 2.0)&(FM, 1.53)\\
\hline 
\end{tabular}
\begin{tablenotes}
  \item[1] \footnotesize $^{1}${Present calculations.} 
  \item[2] $^{2}$Calculations from Ref. \cite{Kulish}. 
	\item[3] $^{3}$Calculations from Ref. \cite{Botana}.
  \end{tablenotes}
\label{GS} 
\vspace{-6 mm}
\end{table}

In general, the ground state in metal dihalides is influenced by either the AFM direct exchange or the FM superexchange, the detailed mechanism of these interactions can be found in Refs. \cite{Goodenough1,Goodenough2,Kanamori1,Kanamori2,Anderson}. These two interactions are determined by the bonding cation-anion-cation where the electron hops from an X site to the nearest X site following Hund's rule and may lead to a spin frustration as explained previously. As for FeI$_{2}$, since the ground state is an FM state, the FM superexchange overcomes the AFM direct exchange. The different situation appears in CoI$_{2}$ and NiI$_{2}$ even though they have the SP ground state. The interaction will be determined by the total energy difference between the FM ($\phi=0$) and AFM ($\phi=1$) states. As shown in Fig. \ref{stiff}(a), the interaction prefers the AFM direct exchange (FM superexchange) for CoI$_{2}$ (NiI$_{2}$) because the AFM state (FM state) is more stable than the FM state (AFM state).   
\begin{figure}[h!]
\vspace{-4mm}
\quad\quad\includegraphics[scale=0.6, width =!, height =!]{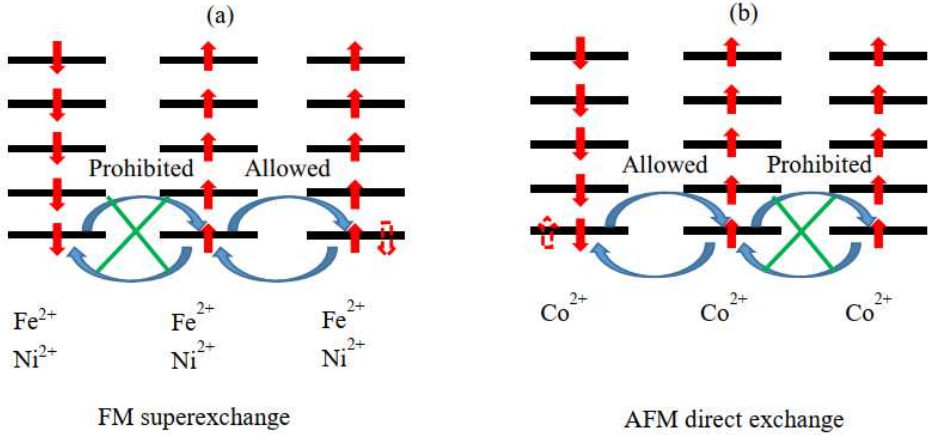}
\vspace{-2mm}
\caption{\label{order} (Color online) Schematic illustration of magnetic orders in FeI$_{2}$ and NiI$_{2}$ (a) and CoI$_{2}$ (b).} 
\end{figure}

Figure \ref{order} shows a simple illustration of how the FM superexchange and AFM direct exchange work due to electron hopping from an X atom to its nearest neighbor X atom in appropriate with Hund's rule. For the FM superexchange, the net magnetic moment of the neighboring X atom (Fe or Ni atom) becomes parallel to reduce the kinetic energy. This activates the inter-site virtual hopping and thus delocalizes the electrons over X-I-X. Conversely, when the kinetic energy reduces because of antiferromagnetically coupled magnetic moments of neighboring X atom (Co atom), it leads to the AFM direct exchange. Furthermore, when a spin competition happens from these two interactions, it may frustrate the magnetic moment to generate an SP ground state, as shown in CoI$_{2}$ and NiI$_{2}$.       	
\begin{figure}[h!]
\vspace{-6mm}
\quad\quad\includegraphics[scale=0.6, width =!, height =!]{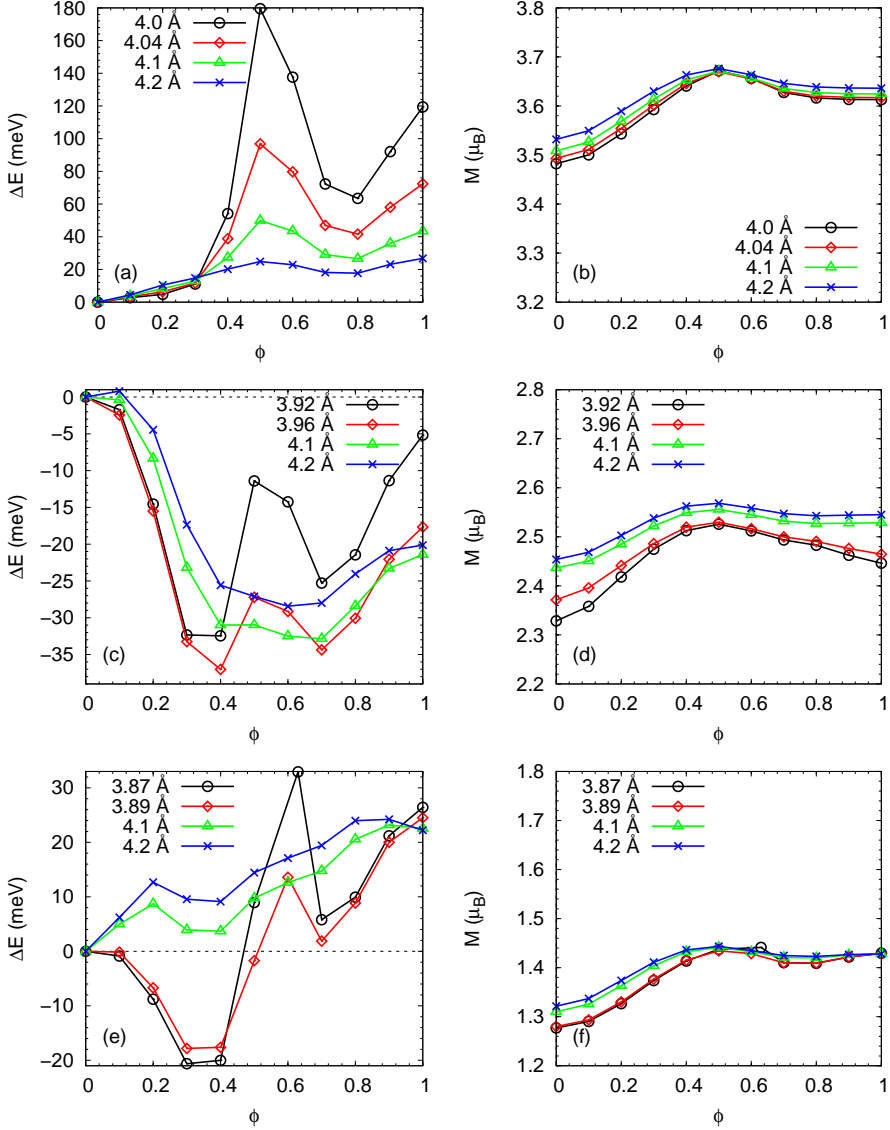}
\vspace{4mm}
\caption{\label{lattice} Dependence of ground state on lattice parameter (a, c, e) and related magnetic moments (b, d, e) for the non-doped case. Here, FeI$_{2}$, CoI$_{2}$ and NiI$_{2}$ are depicted in (a, b), (c, d), and (e, f), respectively. In addition, the total energy difference is defined as $\Delta E=E(\phi)-E(\phi=0)$.}
\end{figure}

Regarding the metal dihalides, our previous studies reported that the ground state of metal dichlorides can be tuned by varying the lattice parameter \cite{Teguh1, Teguh2}. So, our next investigation is to check if the ground state of metal diiodides can also change due to varying the lattice parameter. Figures \ref{lattice}(a, c, e) display the ground state of metal diiodides with respect to the lattice parameter. As for FeI$_{2}$ and CoI$_{2}$, the ground states are robust to the lattice constant. However, we see a change of state from the SP state to the FM state, which means that the ground state of NiI$_{2}$ is sensitive to the lattice parameter. Regarding the magnetic moment in Figs. \ref{lattice}(b, d, f), all the magnetic moments remain stable for all the lattice parameters, indicating that all the systems possess a high spin state. This state happens due to a weak ligand that the energy difference between $t_{2g}$ and $e_{g}$ in the metal atoms is so weak. This implies that the electron will occupy the empty orbital first by following Hund's rule. Note that varying lattice parameter may also lead to a low spin state, as occurred in the fcc iron \cite{Uhl, Mryasov}.
\begin{figure}[h!]
\vspace{-4mm}
\quad\quad\includegraphics[scale=0.6, width =!, height =!]{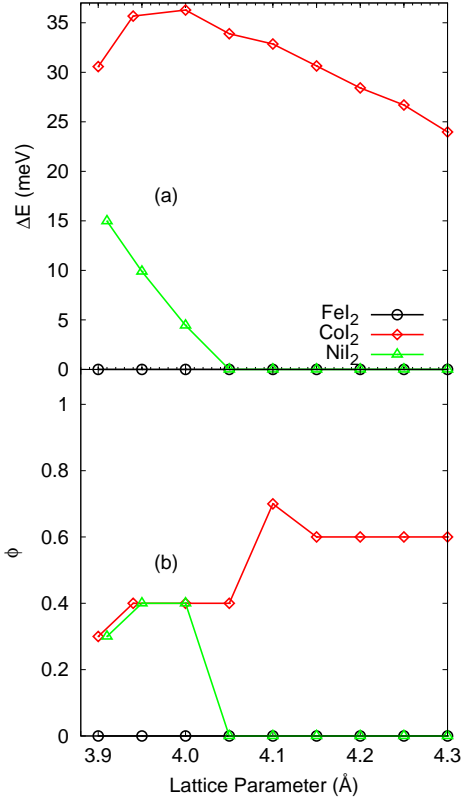}
\vspace{-8mm}
\caption{\label{lattice-dep} Dependence of total energy difference $\Delta E=E(\phi=\phi_{\textrm{lowest}})-E(\phi=0)$ (a) and related ground state $\phi=\phi_{\textrm{lowest}}$ (b).}
\vspace{-4mm}
\end{figure}

A more detailed dependence of the ground state of metal diiodides on the lattice parameter can also be seen in Fig. \ref{lattice-dep}. As for FeI$_{2}$ and CoI$_{2}$, there is no transition of ground state as the lattice parameter increases. This also implies that FeI$_{2}$ is a robust ferromagnet. A different situation occurs in NiI$_{2}$, namely, the FM ground state is preferred at the large lattice parameter. However, for all the lattice parameters, our calculations show that the FM state (AFM state) is more stable than the AFM state (FM state) for FeI$_{2}$ and NiI$_{2}$ (CoI$_{2}$). This indicates that FeI$_{2}$ and NiI$_{2}$ (CoI$_{2}$) are still controlled by the FM superexchange (AFM direct exchange). We will see in the next subsection that this control can be changed by hole-electron doping.
   
\subsection{Doped Case}
\begin{figure}[h!]
\vspace{-6mm}
\quad\quad\includegraphics[scale=0.6, width =!, height =!]{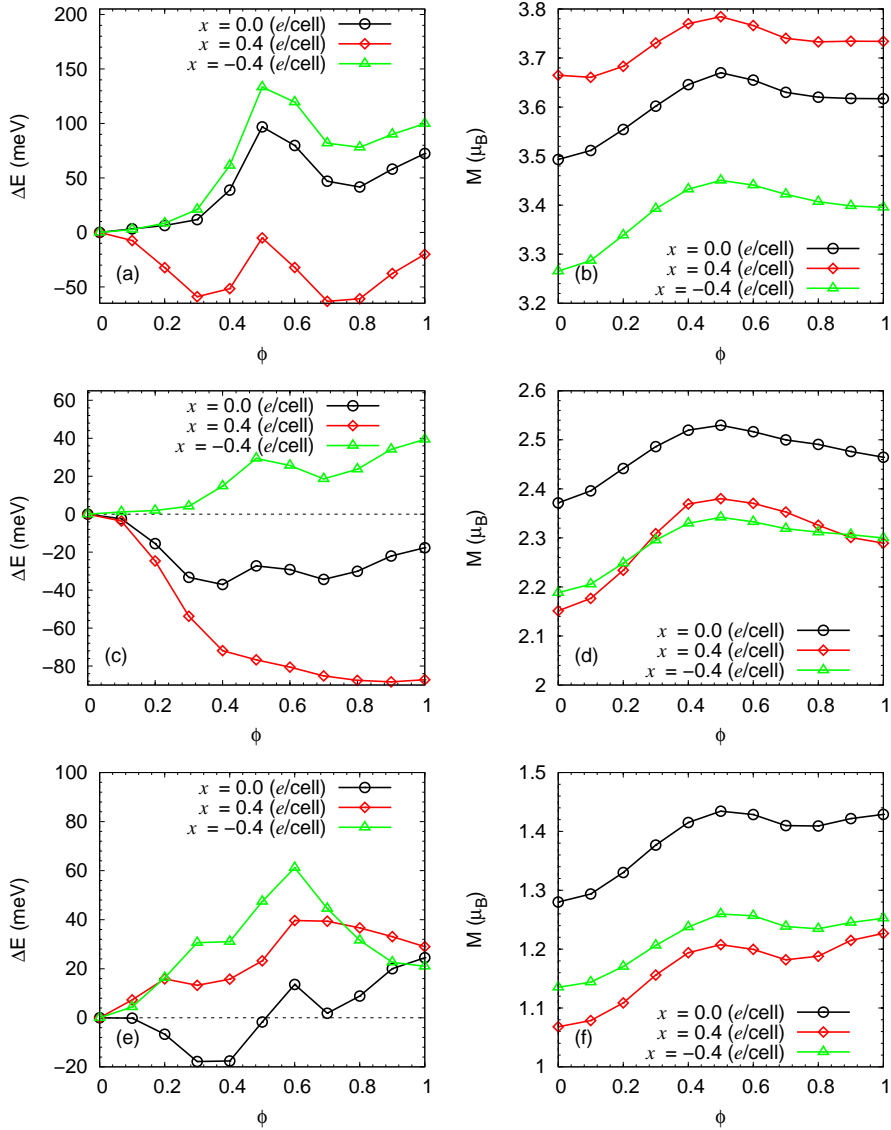}
\vspace{4mm}
\caption{\label{doping} Doping-induced magnetic ground state (a, c, e) and related magnetic moments (b, d, f) for the experimental lattice parameters. We depict FeI$_{2}$, CoI$_{2}$ and NiI$_{2}$ as in (a, b), (c, d), and (e, f), respectively. Here, positive and negative $x$ are addressed to hole and electron doping, respectively. We also define the total energy difference as $\Delta E=E(\phi)-E(\phi=0)$}
\end{figure}

Our final discussion is to consider the magnetic ground state on hole-electron doping. We implement the Fermi level shift approach to run the self-consistent calculations for the doped case \cite{Sawada}. In this approach, a background charge is inserted to make the system neutral. We introduce $x$ as the number of doped concentrations in units of $e$/cell, where the positive (negative) $e$ is addressed to hole (electron) doping. Figures \ref{doping}(a, c, e) show the change of ground state as the doping is introduced. As for FeI$_{2}$, the FM ground state changes to the SP state as $x=0.4$ $e$/cell is applied. Contrarily, the SP ground state in CoI$_{2}$ shifts to the FM ground state when $x=-0.4$ $e$/cell is taken into account. Lastly, the SP ground state in NiI$_{2}$ is replaced by the FM ground state as $x=0.4$ $e$/cell or $x=-0.4$ $e$/cell is introduced. As shown in Figs. \ref{doping}(b, d, f), the magnetic moments tend to stable with respect to the doping, still remaining in the high spin state.
  
\begin{figure}[h!]
\vspace{-6mm}
\quad\quad\includegraphics[scale=0.5, width =!, height =!]{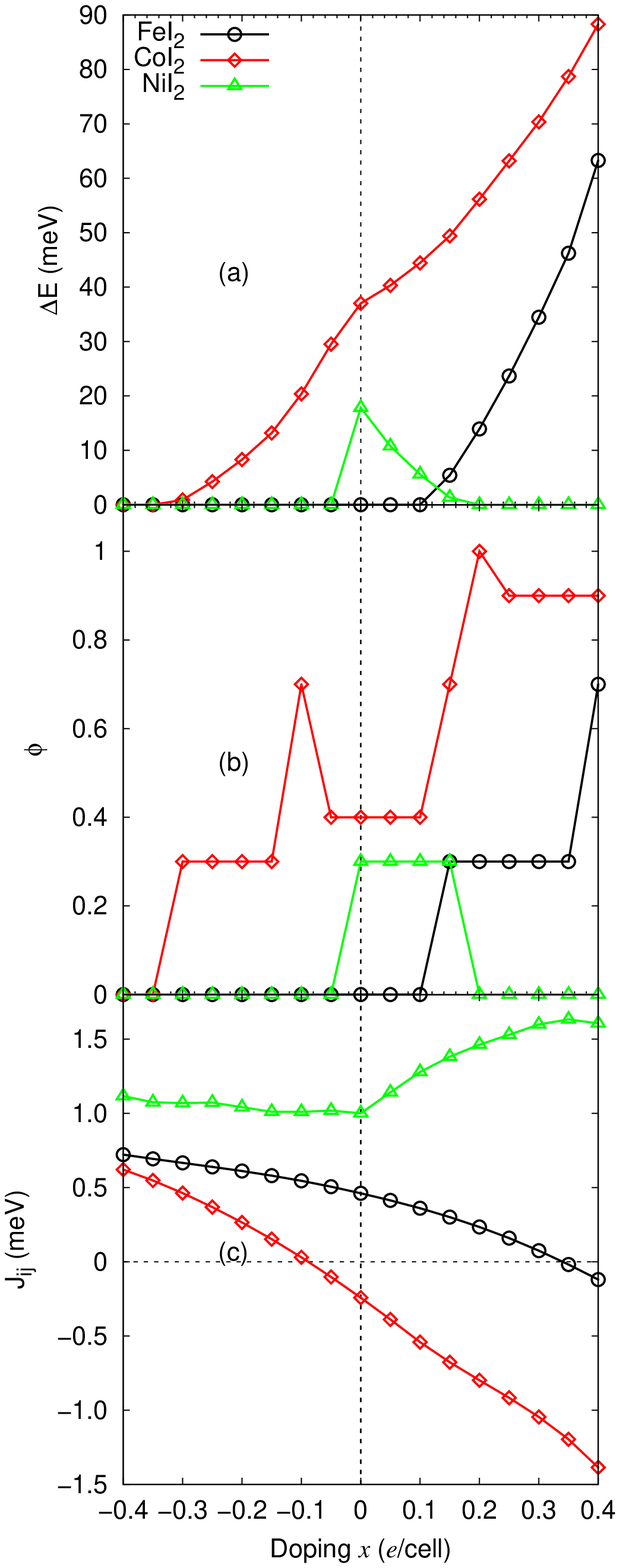}
\vspace{4mm}
\caption{\label{phase}  Magnetic phase transition of XI$_{2}$ (X: Fe, Co, and Ni) with respect to hole-electron doping $x$. Here, we define the total energy difference $\Delta E=E(\phi=0)-E(\phi_{\textrm{lowest}})$ (a) the ground state $\phi=\phi_{\textrm{lowest}}$ (b), and the exchange parameter $J_{ij}$.}
\end{figure}

For the detailed phase transition in the metal diiodides because of the doping, we present Fig. \ref{phase} to see the role of the AFM direct exchange and FM superexchange for creating the phase transition. As shown in Figs. \ref{phase}(a, b), the phase transition arises in the interval of doping $x$. As for FeI$_{2}$, the FM ground state still preserves in the interval of $x \leq 0.1$ $e$/cell while the SP ground state starts to emerge in the interval of $x>0.1$ $e$/cell and no AFM ground state emerges. At the same time, the SP ground state in CoI$_{2}$ still maintains in the range of $-0.3$ $e$/cell $\leq x \leq$ 0.15 $e$/cell and $x \geq 0.25$ $e$/cell, the FM ground state takes the role in the range of $x<-0.3$ $e$/cell, while the AFM ground state only happens in $x=0.2$ $e$/cell. As for NiI$_{2}$, the SP ground state only appears in the interval of $0$ $e$/cell $\leq x \leq$ 0.15 $e$/cell while the FM ground state emerges in the interval of $x<0$ $e$/cell and $x>0.15$ $e$/cell and no AFM ground state emerges. From these results, we conclude that introducing hole doping, in general, frustrates increasingly the spins for all the systems. On the contrary, the frustrated spins due to increasing electron doping only happen in CoI$_{2}$, while the FM ground state becomes robust in FeI$_{2}$ and NiI$_{2}$, thus decreasing the frustrated order to become the FM ground state.

As previously mentioned, the AFM direct exchange competes with the FM superexchange to determine the magnetic order. To discuss the competition, we define the exchange parameter $J_{ij}=(1/12)\Delta E/M^{2}$ \cite{Torun}, which represents the interaction between a magnetic X atom with its nearest neighbor X atom. Here, $\Delta E=E(\phi=1)-E(\phi=0)$ and $M$ mean the total energy difference and the magnetic moment of X atoms. Since an X atom has six nearest neighbor X atoms, we insert 1/12 to prevent the double counting in the calculation. In this case, the positive and negative $J_{ij} $ are addressed to the FM superexchange and AFM direct exchange. As shown in Fig. \ref{phase}(c), we see the different tendencies for all the systems. 

In FeI$_{2}$, the strength of FM superexchange increases (decreases) as the electron (hole) increases, leading to either the FM or SP ground state. In this case, the FM superexchange (AFM direct exchange) works in the range of $x<0.35$ $e$/cell ($x \geq 0.35$ $e$/cell). As for CoI$_{2}$, the strength of AFM direct exchange becomes very strong (weak) as the hole (electron) doping increases, which leads to either the SP or AFM ground state. At this situation, the FM superexchange (AFM direct exchange) performs in the range of $x \leq -0.1$ $e$/cell ($x>-0.1$ $e$/cell). As for NiI$_{2}$, the FM superexchange wins against the AFM direct exchange for both the non-doped and doped cases, which is preferable to the SP or FM ground state. For this case, only the FM superexchange proceeds for all doping.  
\begin{figure}[h!]
\vspace{-4mm}
\quad\quad\includegraphics[scale=0.6, width =!, height =!]{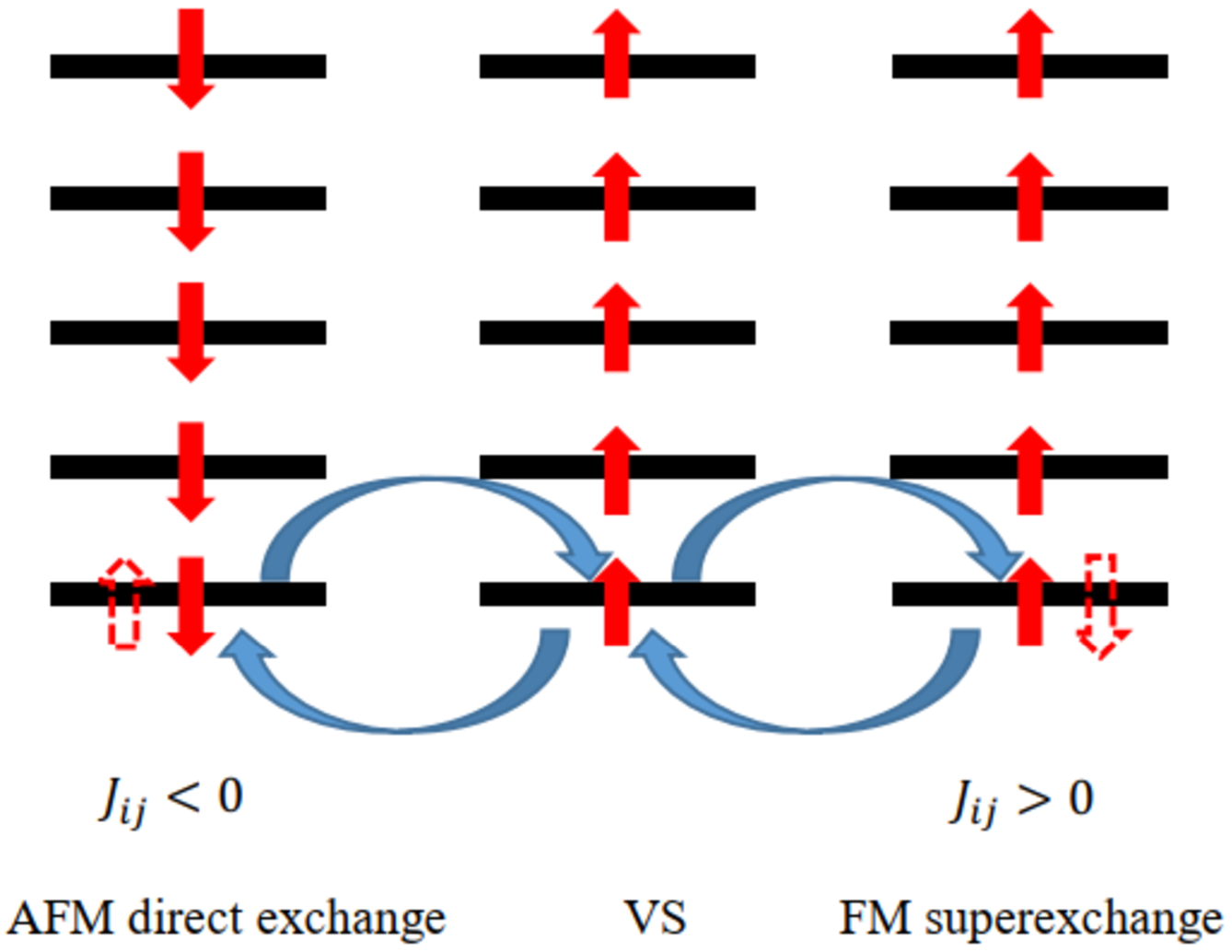}
\vspace{-2mm}
\caption{\label{compete} (Color online) Schematic illustration of competition between AFM direct exchange and FM superexchange.} 
\end{figure}  

As illustrated in Fig. \ref{compete}, introducing the doping lets the Coulomb repulsion to control the occupation of electron in each X site, where the electron concentration can be either reduced by hole doping or increased by electron doping in the 3$d$ orbital. When the FM superexchange wins against the AFM direct exchange ($J_{ij}>0$), the kinetic energy reduces as the magnetic moments of nearest neighbor X atoms become parallel and the electron over X-I-X becomes delocalized. This condition occurs by increasing the electron doping for all the systems as shown in Fig. \ref{phase}(c). On the contrary, when the hole doping increases for FeI$_{2}$ and CoI$_{2}$, the kinetic energy also reduces if the magnetic moments of nearest neighbor X atoms become antiparallel, thus dominated by the AFM direct exchange ($J_{ij}<0$). It also becomes the electron over X-I-X delocalized. As for NiI$_{2}$, the FM superexchange still does exist as hole doping increases, thus indicating a robust ferromagnet with respect to the doping. 
                                                       			
\section{Conclusions} 
In summary, we explore the magnetic ground state of two-dimensional monolayer metal diiodides XI$_{2}$ (X: Fe, Co, and Ni) by utilizing the generalized Bloch theorem (GBT) within the self-consistent flat spin spiral calculations. FeI$_{2}$ exhibits an FM ground state while CoI$_{2}$ and NiI$_{2}$ show an SP ground state. Based on the spin-spin interaction, CoI$_{2}$ prefers the AFM order while NiI$_{2}$ tends to the FM order, even though having the same SP ground state. So, the SP ground state is a manifestation of frustrated spin due to competition between the FM and AFM orders. The emergence of SP state without any external treatments shows that monolayer metal diiodides should be able to generate the multiferroic property that is very useful for applicable spintronic instruments.

We also show that the new ground state can be tuned from the initial ground state by applying hole-electron doping. This utilizes the spin-spin competition between the FM and AFM exchange orders, in which the inter-site virtual electron hopping determines the ground state. The available ground state due to the magnetic configuration of nearest neighbor metal atoms will appear if the kinetic energy can be reduced, thus delocalizing the electron. This mechanism also creates a frustrated spin which leads to an SP ground state. Our findings highlight the importance of introducing the doping to generate a multiferroic property in the metal dihalides for the general case.       
\section*{Acknowledgments}
This research was performed by using a personal high computer and available facilities at Universitas Negeri Jakarta. We hereby state that this research is an independent work without any fundings.

\end{document}